\documentclass[aps,prb,twocolumn,floatfix,showpacs,superscriptaddress]{revtex4-1}
\usepackage{graphicx}
\usepackage{txfonts}
\usepackage[colorlinks]{hyperref}

\begin{document}


\title{Pressure-driven 4$f$ localized-itinerant crossover in heavy fermion compound CeIn$_{3}$: A first-principles many-body perspective}
\author{Haiyan Lu}
\affiliation{Beijing National Laboratory for Condensed Matter Physics, and Institute of Physics, Chinese Academy of Sciences, Beijing 100190, China}
\affiliation{Science and Technology on Surface Physics and Chemistry Laboratory, P.O. Box 9-35, Jiangyou 621908, China}
\author{Li Huang}
\email[Corresponding author. email:]{lihuang.dmft@gmail.com}
\affiliation{Science and Technology on Surface Physics and Chemistry Laboratory, P.O. Box 9-35, Jiangyou 621908, China}
\affiliation{Department of Physics, University of Fribourg, 1700 Fribourg, Switzerland}
\date{\today}


\begin{abstract}
The localized-itinerant nature of Ce-4$f$ valence electrons in heavy fermion compound CeIn$_{3}$ under pressure is studied thoroughly by means of the combination of density functional theory and single-site dynamical mean-field theory. The detailed evolutions of electronic structures of CeIn$_{3}$, including total and partial density of states, momentum-resolved spectral functions, and valence state histograms etc., are calculated in a wide pressure range where the corresponding volume compression $V/V_0 \in [0.6,1.0]$ (here $V_0$ is the experimental crystal volume) at $T \cong 116$ K. Upon increasing pressure, two strong peaks associated with the Ce-$4f$ states emerge near the Fermi level, and the $c$-$f$ hybridization and valence state fluctuation are enhanced remarkably. Moreover, the kinetic and potential energies raise, while the occupancy, total angular momentum, and low-energy scattering rate of the Ce-$4f$ electrons decline with respect to pressure. All the physical observables considered here exhibit prominent kinks or fluctuations in $V/V_0 \in [0.80,0.90]$, which are probably the desired fingerprints for the Ce-4$f$ localized-itinerant crossover.
\end{abstract}


\maketitle


\section{introduction\label{sec:intro}}

The rare-earth elements and their compounds exhibit a variety of fascinating and exotic properties, such as heavy fermions~\cite{coleman:2007,RevModPhys.56.755,PhysRevLett.35.1779}, Dirac fermions~\cite{nc:2013,nasser:2016}, Kondo insulator~\cite{peter:2000}, Topological Kondo insulator~\cite{PhysRevLett.104.106408}, Racah materials~\cite{srep:2015}, and mixed-valence (or valence fluctuation) behaviors~\cite{adroja1991126}, just to name a few. They have attracted a lot of interests and attentions in recent years, and considerable achievements have been achieved~\cite{coleman:2007}. Nowadays it is revealed that the rich physics of rare-earth systems is mainly attributed to the complex characters of their 4$f$ valence electrons and the entanglement between the spin, orbital, and lattice degrees of freedom. Generally speaking, the 4$f$ valence electrons are not only correlated, but also Janus-faced (i.e., sometimes the electrons are localized and inertial, while at other times the electrons are itinerant and participate in the bonding actively) which depends on their surroundings. The dual nature of the 4$f$ valence electrons allows the intriguing properties of rare-earth systems to be easily changed or ``tuned" via some external conditions, such as pressure, temperature, chemistry, and electromagnetic field, etc~\cite{RevModPhys.56.755}.

The 4$f$ localized-itinerant transition or crossover in rare-earth systems is one of the longstanding research topics in the condense matter physics. Typical examples where pressure is used to regulate the 4$f$ states of rare-earth elements from localized to delocalized are Ce, Nd, and Gd~\cite{PhysRevB.72.115125,PhysRevLett.94.036401,PhysRevLett.96.066402}. As the pressure is increased these elements will undergo structural phase transitions companied by enormous volume collapses. Another interesting example where chemical composition plays a pivotal role is about the evolution of 4$f$ valence states in the Ce$T$In$_{5}$ (where $T$ = Co, Rh, and Ir) compounds~\cite{PhysRevLett.108.016402,Shim1615}. The Ce-4$f$ electrons in CeCoIn$_{5}$ and CeIrIn$_{5}$ are itinerant. So they have enlarged electron Fermi surfaces due to the contributions of coherent 4$f$ electrons (guaranteed by the Luttinger theorem). On the contrary, CeRhIn$_{5}$ has localized 4$f$ electrons, so its Fermi surfaces have similar geometry but smaller size. In addition, the temperature effect on the 4$f$ localized-itinerant transition or crossover is an important and very controversially discussed issue as well. Now it is widely believed that upon increasing temperature, the 4$f$ electrons will evolve from itinerant to localized, and manifest themselves by the dramatic change of corresponding physical observables. A natural and straightforward expectation is that the Fermi surfaces will change from large to small volume~\cite{PhysRevLett.108.016402,Shim1615,PhysRevB.84.195141}. However, very recent experiments using angle-resolved photoemission spectroscopy (ARPES)~\cite{PhysRevX.5.011028} demonstrate that the Fermi surface of heavy fermion compound YbRh$_{2}$Si$_{2}$ doesn't change its size or geometry over a considerable temperature range, which is in remarkable contrast to the previously experimental results~\cite{PhysRevB.55.11714,PhysRevLett.106.136401,PhysRevB.77.155128} and theoretical predictions~\cite{PhysRevLett.108.016402,Shim1615,PhysRevB.84.195141}. In order to elucidate this discrepancy, advanced experimental and theoretical investigations are highly imperative. 

Now let's focus on CeIn$_{3}$, which crystallizes in cubic AuCu$_{3}$ structure (space group Pm-3m) and is an important paradigm for heavy fermion and Kondo lattice compounds. The ground state of CeIn$_{3}$ is antiferromagnetic with Ne\'{e}l temperature $T_N = 10$ K~\cite{PhysRevB.22.4379}. Under moderate pressure $P_c \cong 2.65 $ GPa where $T_{N}$ is suppressed to be zero, CeIn$_{3}$ shows unconventional superconductivity below 170 mK~\cite{PhysRevB.65.024425}. It was proposed that the pairing mechanism in the pressure-induced superconducting state is likely related to the magnetic fluctuations~\cite{mathur:1998}, so that CeIn$_{3}$ has been regarded as one of the best tentative compounds to study the interplay between magnetic ordering and superconductivity. Many efforts have been done to investigate the change of its low temperature ($T <T_N$) electronic and magnetic structures under pressure up to 10~GPa via the ARPES, de Haas-van Alphen (dHvA) experiments~\cite{Settai2004223,yoshi:2012,rikio:2005,rikio:2006,PhysRevLett.93.247003,pnas:2009}, $^{115}$In nuclear quadrupole resonance (NQR) spectroscopy~\cite{Thessieu20009,Kohori200012,Kawasaki2003541,PhysRevB.65.020504,PhysRevB.66.054521,PhysRevB.77.064508}, and theoretical calculations~\cite{Ilkhani2008,PhysRevB.80.125131,PhysRevB.65.054405,PhysRevB.75.205130,osamu:2012,Suzuki20081318}, etc. Until now its $P-T$ magnetic phase diagram at low temperature and low pressure regime has been well established. When $T < T_N$, there exists some clues and hints in the experimental results that a 4$f$ localized-itinerant transition probably takes place around 1.0 GPa~\cite{PhysRevB.79.214428,Thessieu20009}. When $T > 15$ K, $T^{*}$ which denotes the characteristic temperature for the localized-itinerant crossover of 4$f$ electrons, exhibits approximately linear behavior with respect to pressure~\cite{PhysRevB.77.064508,Thessieu20009}. We believe that the 4$f$ localized-itinerant transition or crossover in CeIn$_{3}$ will be a key factor to understand its complicated electronic structures and related physical properties. However, this issue is less concerned in the available theoretical investigations. In particular, the changes of its electronic and magnetic structures in the transition have not been touched too much.

In order to fill in this gap, we endeavored to unveil the evolution of electronic structures of CeIn$_{3}$ under pressure by using a first-principles many-body approach, specifically, the density functional theory merged with the single-site dynamical mean-field theory (dubbed as DFT + DMFT)~\cite{RevModPhys.78.865,RevModPhys.68.13}. The purpose of the present work is thus three-folds: (1) Further examine the availability of the DFT + DMFT method in the simulations of strongly correlated heavy fermion compounds. Here CeIn$_{3}$ is taken as a prototype for benchmark. (2) For a given temperature $T^{*}$, try to predict the critical pressure $P_c$ where the $4f$ localized-itinerant transition or crossover in CeIn$_3$ may occur. (3) Explore the subtle changes of electronic structures in this transition or crossover and elucidate the underlying physics. These calculated results can serve as critical predictions, and enrich our understanding about the pressure-driven electronic phase transition in CeIn$_{3}$.

The rest of this paper is organized as follows: Section~\ref{sec:method} introduces the technical details for the DFT and DMFT calculations. Particularly, the construction for the DFT + DMFT Hamiltonian, and the strategies to accelerate the DMFT calculations are explained. In Sec.~\ref{sec:results} the main results, involving the physical properties of CeIn$_{3}$ at ambient pressure and the evolution of electronic structures under pressure are presented and discussed in details. In this section, we pay special attention to the redistribution of valence state histograms during the itinerant-localized transition or crossover and analyze its consequences on occupancy, total angular momentum, kinetic and potential energies, etc. Finally, a brief conclusion is given in Sec.~\ref{sec:summary}.  

\section{method\label{sec:method}}

Since the Ce-4$f$ states in CeIn$_{3}$ is undoubtedly correlated~\cite{osamu:2012,PhysRevB.75.205130,Suzuki20081318}, the strong Coulomb interaction among the Ce-4$f$ electrons has to be taken into consideration carefully to obtain a reasonable description of their localized-itinerant transition. On the other hand, the DFT + DMFT approach, which combines the first-principles aspect of DFT with the non-perturbative many-body treatment of local interaction effects in DMFT, is probably the most powerful established method to study the electronic structures of strongly correlated materials~\cite{RevModPhys.78.865,RevModPhys.68.13}. It has been successfully applied in the studies of many heavy fermions or rare-earth systems, such as the $\alpha-\gamma$ phase transition in Ce~\cite{PhysRevLett.94.036401,PhysRevLett.96.066402,PhysRevB.72.115125}, pressure-driven valence fluctuation in Yb~\cite{PhysRevLett.102.246401}, atomic multiplets of mixed-valence compound SmB$_{6}$~\cite{srep:2015}, and temperature-dependent localized-itinerant transition in CeIrIn${_5}$~\cite{PhysRevLett.108.016402,Shim1615}, etc. Inspired by the previous achievements, in the present work we adopted the DFT + DMFT method to perform charge fully self-consistent calculations to explore the fine electronic structures of CeIn$_{3}$ as a function of hydrostatic pressure.  

The DFT + DMFT iterations can be split into two individual parts, DFT and DMFT. The main task of DFT part is to generate the Kohn-Sham single-particle Hamiltonian $\hat{H}_{\text{KS}}$. Here we used the \texttt{WIEN2K} code~\cite{wien2k}, which implements the full-potential linear augment plane wave formalism to accomplish the DFT calculations. The cutoff parameters satisfied $R_{\text{MT}}K_{\text{MAX}} = 7.0$, the Brillouin zone integration was done on a $17 \times 17 \times 17$ Monkhorst-Pack $k$-points, and the spin-orbital coupling (SOC) was also included. We reduced the lattice constants of CeIn$_{3}$ gradually to simulate the effect of pressure, while the cubic symmetry was kept all the time. In the DMFT part, the Hamiltonian obtained from DFT was supplemented with a Coulomb interaction term $\hat{H}_{\text{int}}$ for the Ce-4$f$ orbitals and a double counting term for self-energy function $\Sigma_{\text{dc}}$, and then the resulting multi-orbital lattice model
\begin{equation}
\label{eq:ham}
\hat{H}_{\text{DFT+DMFT}} = \hat{H}_{\text{KS}} + \hat{H}_{\text{int}} - \Sigma_{\text{dc}},
\end{equation}
 was solved utilizing the DMFT method~\cite{RevModPhys.78.865,RevModPhys.68.13}. We chose the \texttt{DMFT\_W2K} code developed by K. Haule \emph{et al.}~\cite{PhysRevB.81.195107} to carry out this task. The $\hat{H}_{\text{int}}$ term is parameterized with the Slater integrals $F^0$, $F^2$, $F^4$ and $F^6$. For the $4f$ electronic systems, the following relations are applied~\cite{PhysRevB.59.9903}:
\begin{equation}
U=F^0,\ J=\frac{2}{45}F^2+\frac{1}{33}F^4+\frac{50}{1287}F^6,
\end{equation}
and
\begin{equation}
F^4=\frac{451}{675}F^2,\ F^6=\frac{1001}{2025}F^2,
\end{equation}
where $U$ is the Coulomb interaction strength and $J$ the Hund's exchange parameter. Here we adopted $U=6.2$\ eV and $J=0.7$\ eV, which were exactly accordance with the values reported in the literatures~\cite{Ilkhani2008,PhysRevB.75.205130}. As for the double counting term $\Sigma_{\text{dc}}$, the fully localized limit (FLL) scheme was used~\cite{jpcm:1997},
\begin{equation}
\label{eq:dc}
\Sigma_{\text{dc}} = U\Big(N_{f}-\frac{1}{2}\Big) - \frac{J}{2} \Big(N_{f}-1\Big),
\end{equation}
where the 4$f$ occupancy $N_f$ was updated dynamically during the DFT + DMFT iterations ($N_f = 1$ in the first iteration). The hybridization expansion version of continuous-time quantum Monte Carlo (dubbed as CT-HYB) impurity solver~\cite{PhysRevLett.97.076405,RevModPhys.83.349,PhysRevB.75.155113} was employed to solve the multi-orbital lattice model as defined in Eq.~(\ref{eq:ham}). We not only used the good quantum numbers $N$ and $J_z$ to reduce the sizes of matrix blocks of the local Hamiltonian, but also truncated the local Hilbert space and just kept the atomic eigenstates with $N \in$ [0,4]. In order to further accelerate the Monte Carlo samplings, the lazy trace evaluation trick~\cite{PhysRevB.90.075149} was applied as well. For each DMFT iteration, $16 \times10^8$ Monte Carlo steps were performed to reach sufficiently high accuracy. 

We performed charge fully self-consistent DFT + DMFT calculations. In most cases, 40 DFT + DMFT iterations are adequate to obtain a well converged charge density $\rho$ and total energy $E_{\text{tot}}$. For each DFT + DMFT iteration, 20 DFT internal cycles and a one-shot DMFT calculation were executed. All of the calculations were carried out at the inverse temperature $\beta=100$ ($T \approx 116$\ K), which is much higher than $T_N$~\cite{mathur:1998}, so it is reasonable to consider only the paramagnetic solutions in the calculations.

\section{results and discussions\label{sec:results}}
\subsection{CeIn$_{3}$ at ambient pressure}

\begin{figure}[t]
\centering
\includegraphics[width=\columnwidth]{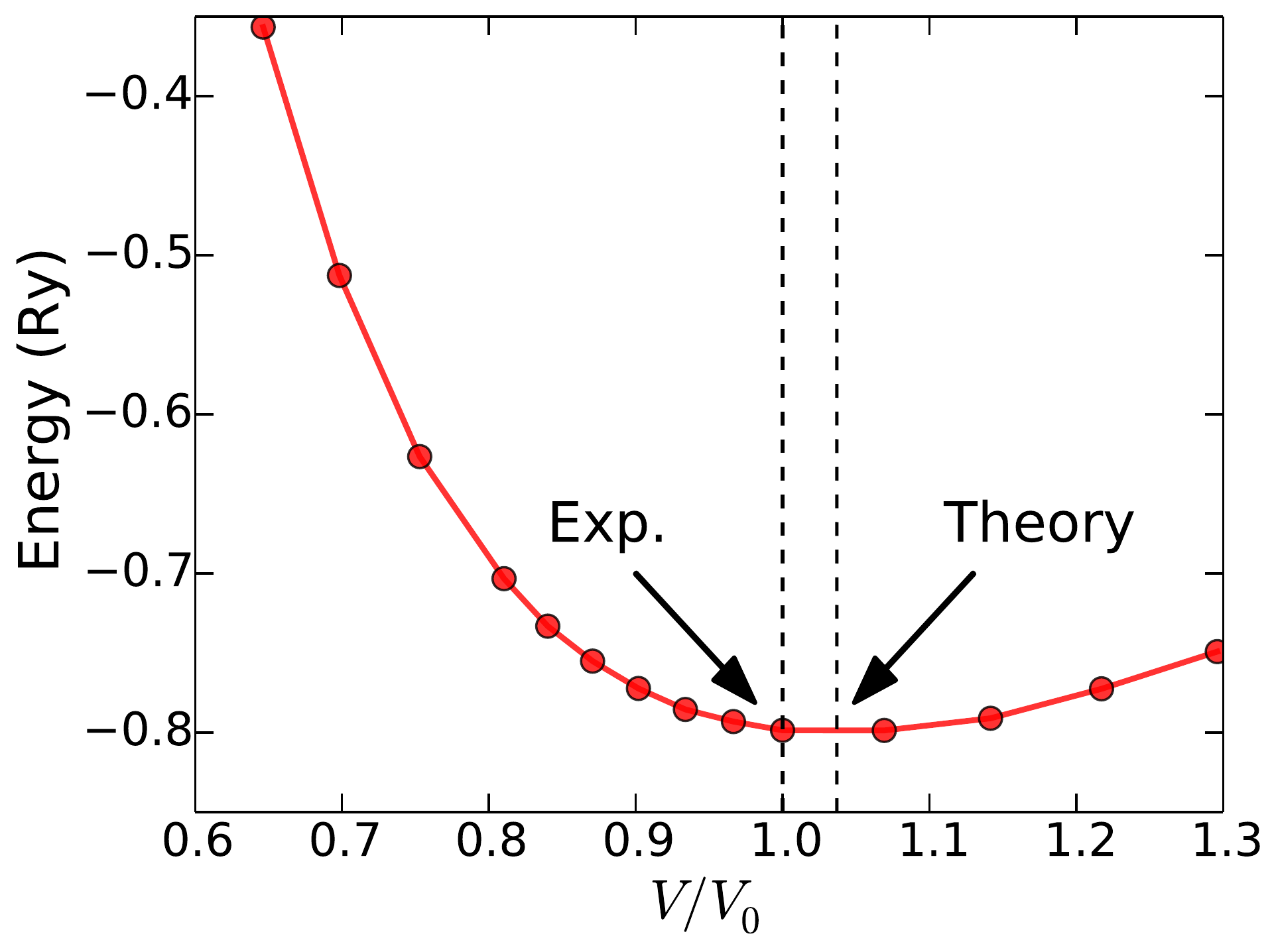}
\caption{(Color online). Calculated $E-V$ relation of CeIn$_{3}$. A reference energy (-53030 Ry) is subtracted from the total energy data. Here and in the following, $V/V_0$ means the volume compression ($V$: current crystal volume, $V_0$: experimental crystal volume~\cite{BENOIT1980293}). The left and right vertical lines mean the experimental and equilibrium volume volumes, respectively. \label{fig:tev}}
\end{figure}

\begin{figure*}[t]
\centering
\includegraphics[width=\textwidth]{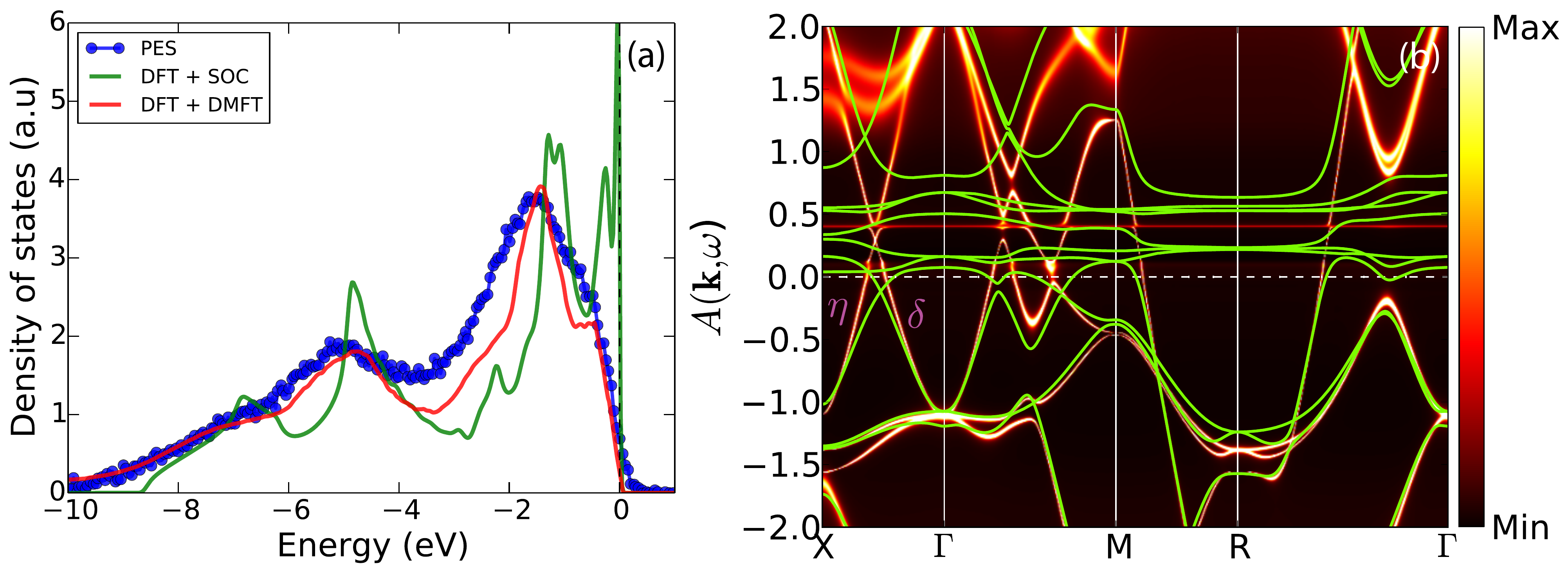}
\caption{(Color online). Electronic structures of CeIn$_{3}$ under ambient pressure ($V/V_0 = 1.00$). (a) Total density of states $A(\omega)$. The experimental data are taken from Reference~[\onlinecite{gam:2008}] and rescaled for a better visualization. The DFT + DMFT and DFT + SOC density of states are firstly multiplied by the Fermi-Dirac function ($\beta = 100.0$), and then smoothed. (b) Momentum-resolved spectral function $A(\mathbf{k},\omega)$ along the high-symmetry lines in the Brillouin zone obtained by DFT + DMFT calculations. The DFT + SOC band structures are represented as bold green lines. The vertical and horizontal lines in (a) and (b) denote the Fermi level, respectively. \label{fig:texp}}
\end{figure*}

It is very important to examine whether the bulk properties and electronic structures of CeIn$_{3}$ under ambient pressure can be correctly reproduced by the DFT + DMFT method before we could apply it to study the pressure-driven 4$f$ localized-itinerant transition or crossover.

At first, we calculated the total energies of CeIn$_{3}$ at different volumes to get the $E(V)$ relation. In the framework of DFT + DMFT, the expression of total energy is given as follows:
\begin{equation}
E_{\text{tot}} = E_{\text{DFT}} + E_{\text{KS}} + E_{\text{pot}} - E_{\text{dc}}.
\end{equation}
Here $E_{\text{DFT}}$ denotes the DFT energy, $E_{\text{KS}}$ the Kohn-Sham band energy correction due to the DMFT density matrix, $E_{\text{pot}}$ the DMFT potential energy, and $E_{\text{dc}}$ the double counting correction~\cite{RevModPhys.78.865,PhysRevB.81.195107}. Actually, the total energy $E_{\text{tot}}$ only depends on the charge density $\rho(\mathbf{r})$ and Matsubara Green's function $G(i\omega_n)$. Once the $E-V$ curve was ready (see Fig.~\ref{fig:tev}), we then used the Birch-Murnaghan equation of states (EOS) to fit it. After that, the bulk parameters, including the bulk modulus $B$ and equilibrium lattice constant $a_0$ were extracted. The calculated $B$ and $a_0$ are 51 GPa and 4.745 \AA, respectively, which are in roughly consistent with the available experimental data 67 GPa~\cite{Oomi1999220} and 4.689 \AA~\cite{BENOIT1980293}.

Second, we accomplished DFT + DMFT calculation using the experimental crystal structure. The fully converged Matsubara self-energy $\Sigma(i\omega_n)$ was firstly converted into its counterpart on real frequency $\omega$ by the maximum entropy method~\cite{jarrell}. Then using $\Sigma(\omega)$ as an essential input, the total density of states $A(\omega)$ and momentum-resolved spectral function $A(\mathbf{k},\omega)$ were evaluated as follows:
\begin{equation}
A(\omega) = \int_{\Omega} d \mathbf{k} A(\mathbf{k},\omega),
\end{equation}
and
\begin{equation}
A(\mathbf{k},\omega) = -\frac{1}{\pi}\Im\frac{1}{\omega + \mu - \hat{H}_{\text{KS}}(\mathbf{k}) - [ \Sigma(\omega) - \Sigma_{\text{dc}}]},
\end{equation}
where $\mu$ is the chemical potential. The calculated results, together with the available PES data~\cite{gam:2008}, are shown in Fig.~\ref{fig:texp}. Note that in order to make a meaningful comparison with the experimental data, the obtained $A(\omega)$ has to be multiplied by a Fermi-Dirac function at first, and then broadened using a Gaussian-like function with suitable smearing parameter $\sigma$. As is clearly seen in Fig.~\ref{fig:texp}(a), the density of states $A(\omega)$ agrees quite well with the PES data. The peaks at $\omega \sim -1.8$ and -5.0 eV, and the shoulder feature at $\omega \sim -0.3$ eV are well reproduced. The $A(\omega)$ obtained by the DFT + SOC method is also shown in this figure as a supplement. Obviously, the peak positions aren't very precise. There is a sharp peak in the vicinity of the Fermi level, which is contrary to the experiment. In Fig.~\ref{fig:texp}(b), the $A(\mathbf{k},\omega)$, together with the band structures from DFT + SOC calculation are shown. In the region below -1.0 eV, there is still a good correspondence between them. However, near the Fermi level (between -1.0 eV and 1.0 eV), there are substantial discrepancies between them. Firstly, the DFT + DMFT bands are strongly renormalized relative to the DFT + SOC bands. Second, there is a flat-band feature accompanied with strong $c-f$ hybridization in the DFT + DMFT bands at $\omega \sim 0.4$ eV, which is associated with the Ce-4$f$ $j = 7/2$ states, while this feature is absent in the DFT + SOC bands. Third, there are some band-crossing structures near -0.2 eV (along the $X$ - $\Gamma$ - $M$ lines) in the DFT + SOC bands, however, in the DFT + DMFT spectra these features are shifted upward to the above of the Fermi level. Very recently, Zhang \emph{et al}.~\cite{yun:2015} conducted the ARPES experiment for the paramagnetic CeIn$_{3}$ at 13 K. Both the experimental valence band structures and the corresponding momentum distribution curves support our results. For example, along the $X$ - $\Gamma$ line the electron-like band $\delta$ and hole-like band $\eta$ only adjoin, instead of intersecting below the Fermi level [see Fig.~\ref{fig:texp}(b)].  

It seems that the failure of the DFT + SOC method in CeIn$_{3}$ is due to the neglect of the 4$f$ electronic correlation effect~\cite{osamu:2012,PhysRevB.75.205130}. However, it could be correctly captured by the DFT + DMFT method. From what has been discussed above, we may safely draw the conclusion that the DFT + DMFT method is a reliable tool to describe the electronic states of CeIn$_{3}$.
 
\subsection{Electronic band structures}
\begin{figure*}[t]
\centering
\includegraphics[width=\textwidth]{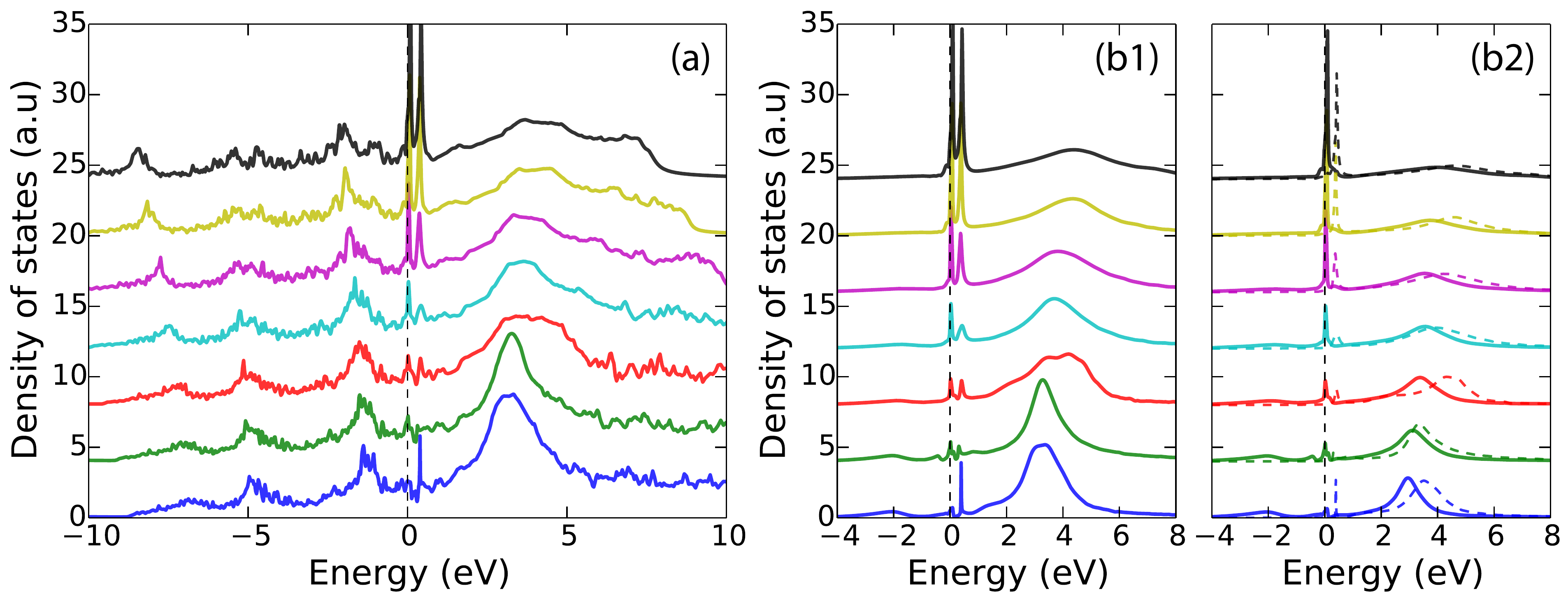}
\caption{(Color online). Evolution of total and partial density of states upon increasing pressure obtained by DFT + DMFT calculations ($\beta$ = 100.0). (a) Total density of states $A(\omega)$. (b1) Ce-4$f$ density of states $A_{4f}(\omega)$. (b2) Partial density of states for the Ce-$4f$ $j$ = 5/2 and 7/2 states. Here $A^{j = 5/2}_{4f}(\omega)$ and $A^{j=7/2}_{4f}(\omega)$ are represented using solid and dashed lines, respectively. $V/V_0$ = 1.00, 0.93, 0.87, 0.81, 0.75, 0.70, and 0.65 (from bottom to top). \label{fig:tdos}}
\end{figure*}

\begin{figure}[t]
\centering
\includegraphics[width=\columnwidth]{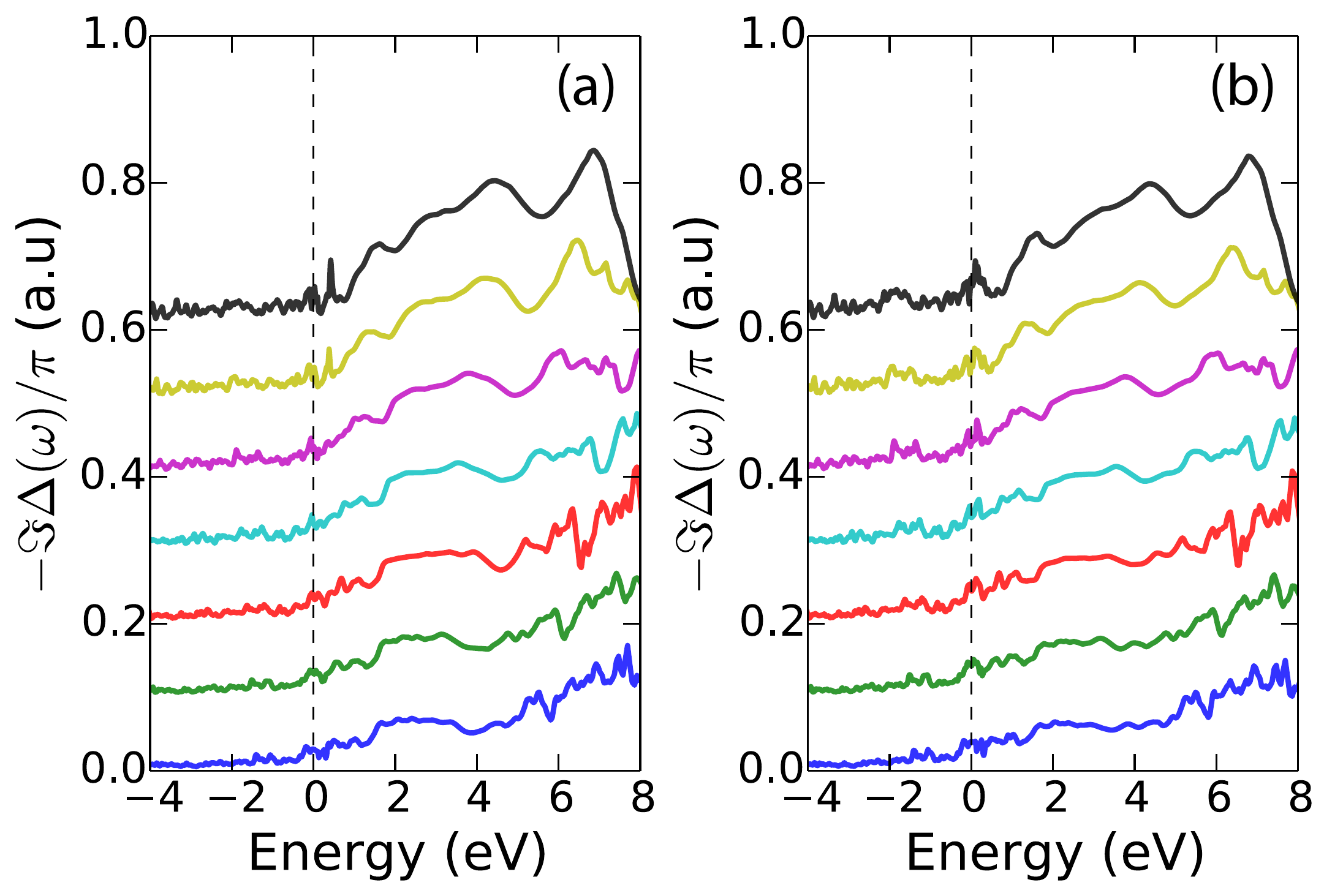}
\caption{(Color online). Evolution of impurity hybridization function $\tilde{\Delta}(\omega)$ upon increasing pressure obtained by DFT + DMFT calculations ($\beta$ = 100.0). (a) For the Ce-4$f$ $j$ = 5/2 states. (b) For the Ce-$4f$ $j$ = 7/2 states. $V/V_0$ = 1.00, 0.93, 0.87, 0.81, 0.75, 0.70, and 0.65 (from bottom to top). \label{fig:thyb}}
\end{figure}

\begin{figure*}[t]
\centering
\includegraphics[width=\textwidth]{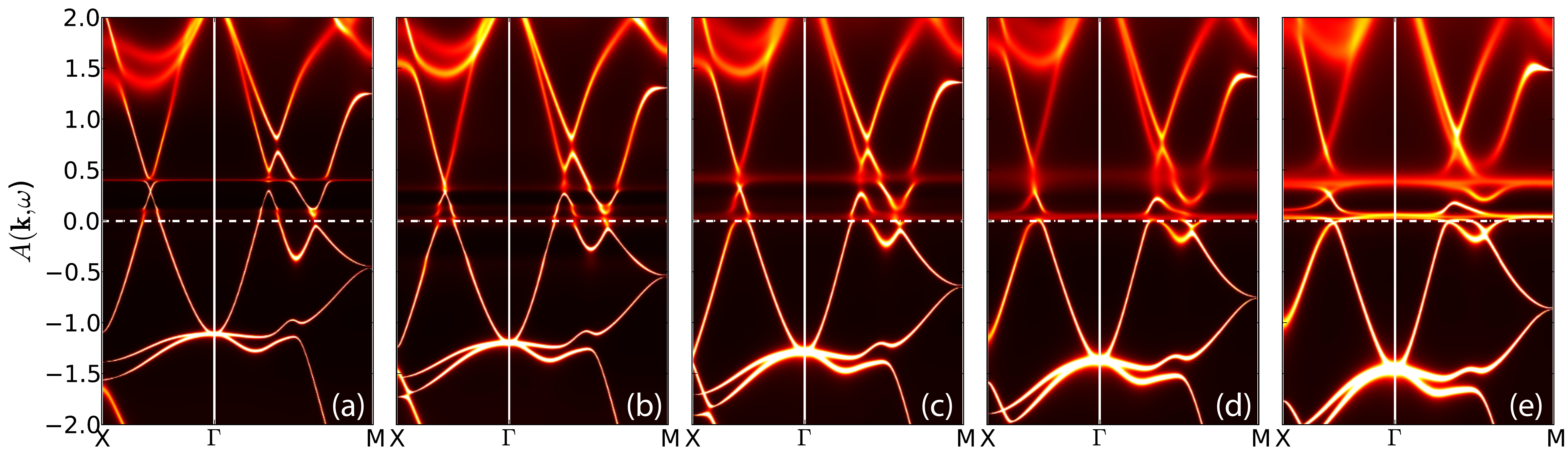}
\caption{(Color online). Evolution of momentum-resolved spectral functions $A(\mathbf{k},\omega)$ upon increasing pressure obtained by DFT + DMFT calculations ($\beta$ = 100.0). $V/V_{0} = $ 1.00 (a), 0.97 (b), 0.87 (c), 0.81 (d), and 0.75 (e). The horizontal dashed lines mean the Fermi level. \label{fig:takw}}
\end{figure*}

Next, we concentrate our attention to the evolution of electronic states of CeIn$_{3}$ upon increasing pressure (or equivalently decreasing volume). In Fig.~\ref{fig:tdos}, the pressure-dependent total and partial (Ce-4$f$) density of states are shown. When the pressure is small, the spectral weight of $A_{4f}(\omega)$ in the vicinity of the Fermi level is nearly trivial, which indicates at that time the 4$f$ electrons in CeIn$_{3}$ are almost completely localized and have little contributions to the chemical bonding. As the pressure is increased, the coherent peak grows up quickly. Especially, when $V/V_0 = 0.87$, its spectral weight becomes considerable. So as a rough estimation, we speculate that the Ce-4$f$ localized-itinerant crossover in CeIn$_{3}$ takes places around $V/V_0 \sim 0.87$. Due to the SOC effect, the Ce-4$f$ orbitals are split into the $j = 5/2$ and $j = 7/2$ states~\cite{cfs}. The peaks located at $\omega \sim $ 0.1 eV and 0.4 eV are mainly associated with the $j = 5/2$ and $j = 7/2$ states, respectively. The distance between the two peaks is about 0.3 eV, which is approximately equal to the spin-orbit splitting $\Delta_{\text{SO}} = 280$ meV~\cite{PhysRevB.56.1620}. Apparently, the pressure-driven localized-itinerant crossover in CeIn$_{3}$ is dominated by the low-lying $j = 5/2$ states, whose spectral weights are strongly heightened by the pressure. The spectral weight of the high-lying $j = 7/2$ states is also enhanced with respect to the pressure, but the change is less intense. The lower and upper Hubbard bands of the Ce-$4f$ orbitals are approximately located from $-4$ to $-1$ eV and $1$ to $8$ eV, respectively. When the pressure is increased the lower Hubbard bands are suppressed and finally smeared out, which is related to the gradual decrease of 4$f$ occupancy. On the other hand, the upper Hubbard bands broaden out, and shift to higher frequency saliently. As a consequence, the hybridization between them and the conducting bands becomes stronger and stronger at the same time. In order to clarify this problem, we further evaluated the impurity hybridization function $\tilde{\Delta}(\omega)$ for the Ce-4$f$ orbitals:
\begin{equation}
\tilde{\Delta}(\omega) = -\frac{\Im \Delta(\omega)}{\pi}.
\end{equation}
The obtained results are depicted in Fig.~\ref{fig:thyb}. Clearly, for both the $j = 5/2$ and $j = 7/2$ states, the impurity hybridization functions are enhanced at $\omega \in $[0~eV,6~eV] under pressure, which signals the increment of $c$-$f$ hybridization strength.      

Now let's take a closer look at the evolution of the momentum-resolved spectral functions $A(\mathbf{k},\omega)$ under pressure. Some representative results are shown in Fig.~\ref{fig:takw}. The low-energy band structures of CeIn$_{3}$ are modified by the pressure apparently. The parallelly flat-band structures at $\omega \sim$ 0.1 eV and 0.4 eV, manifested themselves as sharp peaks in the Ce-4$f$ density of states $A_{4f}(\omega)$, are attributed to the contributions from the $j = 5/2$ and $j = 7/2$ states, respectively. Their intensities are very sensitive to pressure or crystal volume. When the pressure is small, the feature for the $j = 5/2$ states is hardly detectable and that for the $j = 7/2$ states is dim. When $V/V_0 \sim 0.87$, the $j = 5/2$ character is visible, while that for the $j = 7/2$ states is broadened. If we continue to shrink the crystal volume, both of them finally become luminous. As for the high-energy part, a set of broad and dispersive bands are seen. They can be interpreted as the Ce- or In-$spd$ bands which leave an imprint in the Ce-$4f$'s lower and upper Hubbard bands via hybridization. These bands are slightly shifted outward and broadened under pressure. 

\subsection{Valence fluctuations}

\begin{figure*}[t]
\centering
\includegraphics[width=\textwidth]{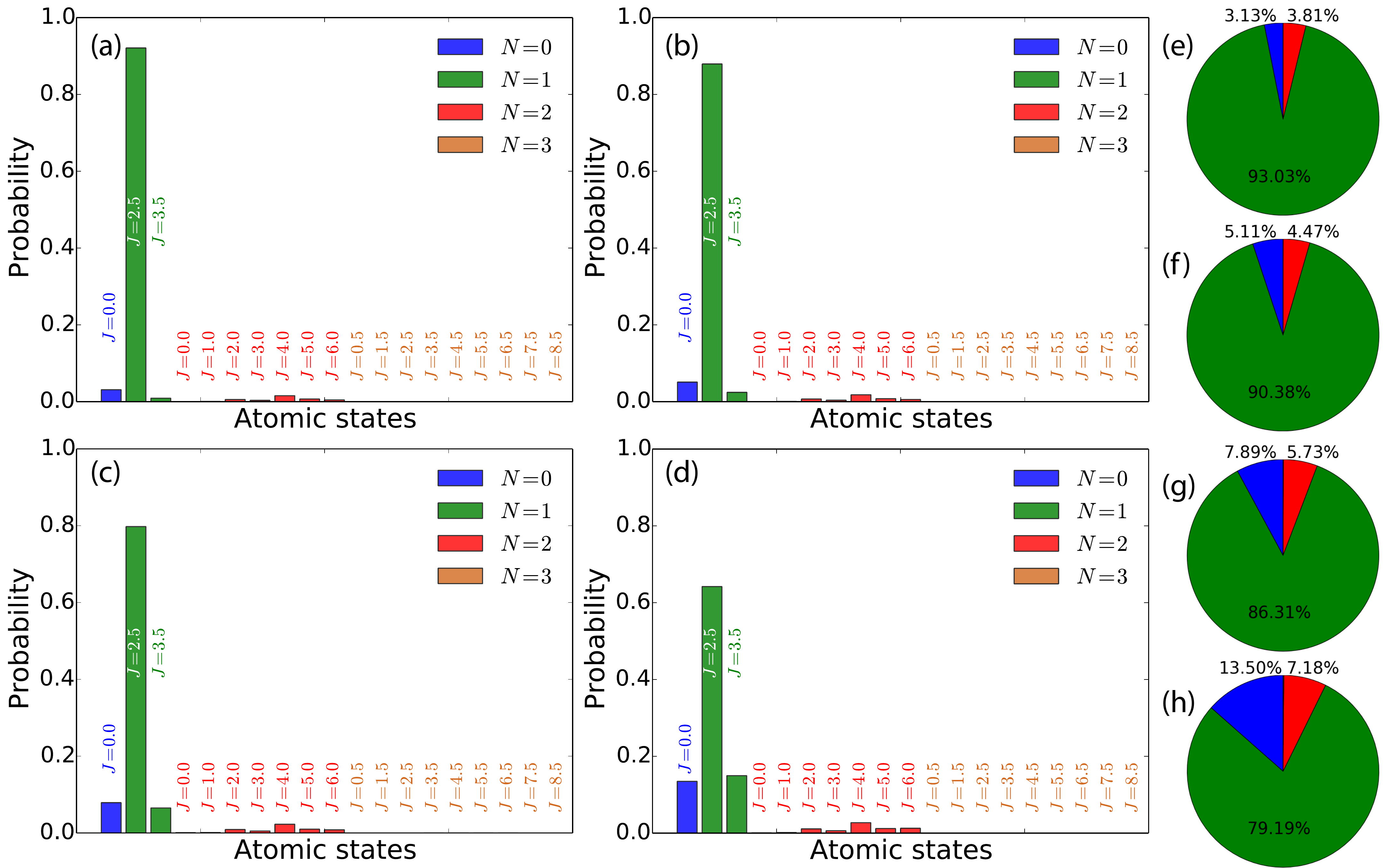}
\caption{(Color online). Distributions of atomic state probability for Ce-4$f$ states at $V/V_{0}$ = 1.00 (a), 0.87 (b), 0.75 (c), and 0.65 (d) obtained by DFT + DMFT calculations ($\beta$ = 100.0). The corresponding distributions of 4$f$ occupancy are summarised in panels (e), (f), (g), and (h), respectively. Note that the percentages for the $N = 3$ atomic states are too small ($< 1\%$) to be seen in (e)-(h). \label{fig:tprob1}}
\end{figure*}

\begin{figure*}[t]
\centering
\includegraphics[width=\textwidth]{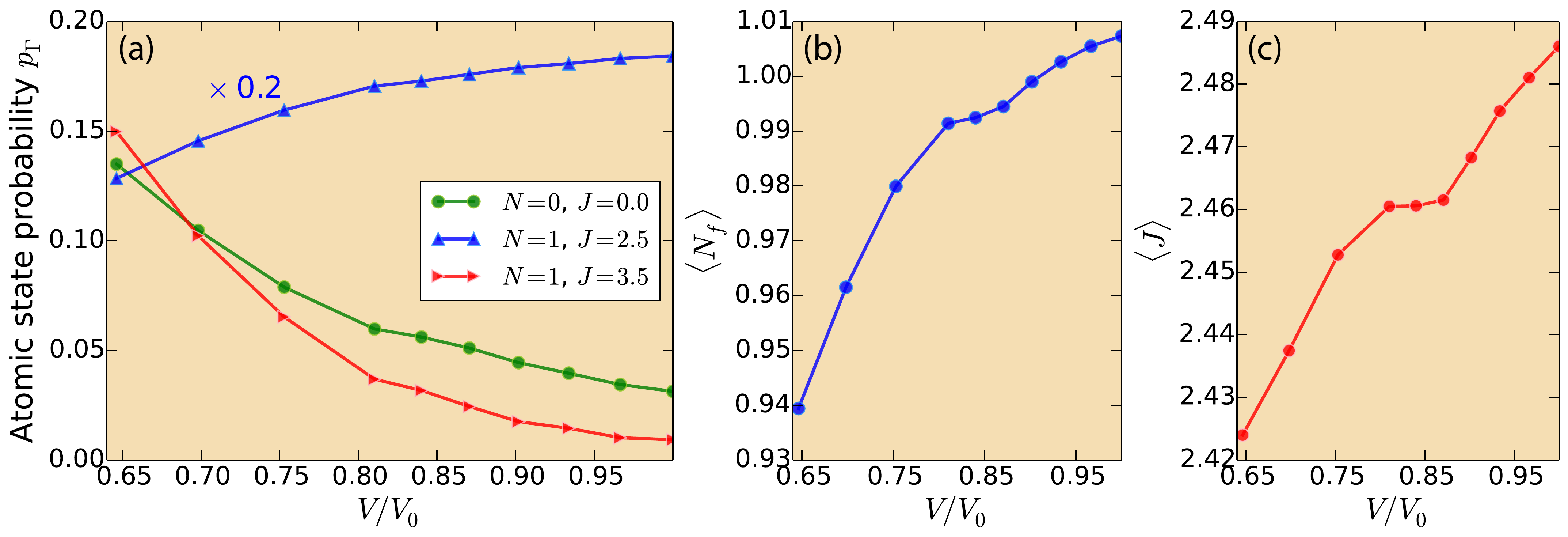}
\caption{(Color online). Evolution of some physical observables upon increasing pressure obtained by DFT + DMFT calculations ($\beta$ = 100.0). (a) Probabilities of representative atomic eigenstates. Here the data for the $|N = 1, J = 2.5\rangle$ state are rescaled (multiplied by a factor of 0.2) for a better visualization. (b) Expected values of Ce-4$f$ occupancy $\langle N_{f} \rangle$. (c) Expected values of Ce-4$f$ total angular momentum $\langle J \rangle$. \label{fig:tprob2}}
\end{figure*}

\begin{figure*}[t]
\centering
\includegraphics[width=\textwidth]{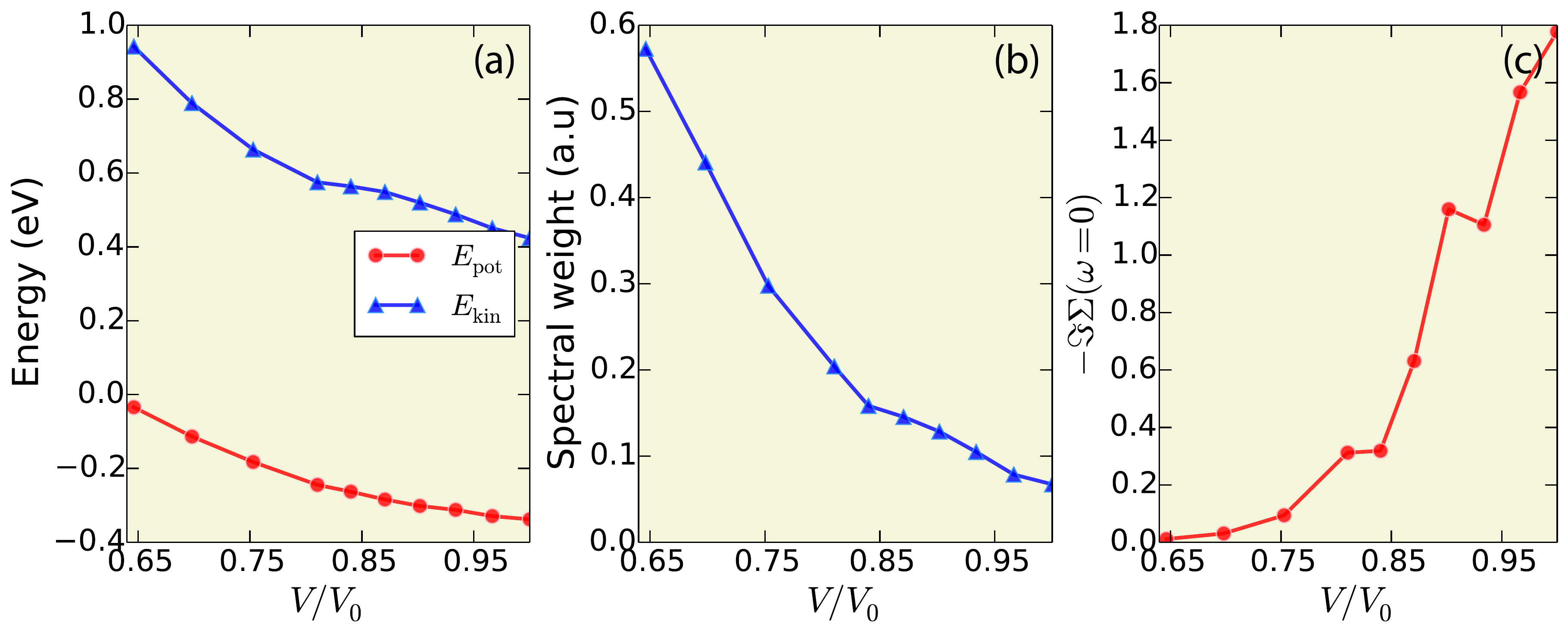}
\caption{(Color online). Evolution of some physical observables upon increasing pressure obtained by DFT + DMFT calculations ($\beta$ = 100.0). (a) Ce-4$f$ potential and kinetic energies, $E_{\text{pot}}$ and $E_{\text{kin}}$. (b) Spectral weight of characteristic peak of Ce-4$f$ $j = 5/2$ states near the Fermi level, $W_{j = 5/2}$. See Eq.~(\ref{eq:w52}) for its explicit definition. (c) Low energy scattering rate $\gamma$ [$\equiv -\Im \Sigma (\omega = 0)$] for the Ce-4$f$ $j = 5/2$ states. \label{fig:tene}}
\end{figure*}

In this subsection, we will focus on the Ce-4$f$ valence fluctuation and the correspondingly physical consequences for the localized-itinerant crossover in CeIn$_{3}$, which has been rarely concerned to our knowledge in the previous studies. 

The best way to analyze the valence fluctuation quantitatively is via the valence state histogram (or atomic state probability), which represents the probability to find a valence electron in a given atomic eigenstate. In other words, the valence state histogram can be considered as lifetime for a given atomic eigenstate. It provides additional information about the dual nature of the Ce-4$f$ electrons. Note that the valence state histogram is a direct output of the CT-HYB impurity solver~\cite{PhysRevB.75.155113}. In Fig.~\ref{fig:tprob1}(a)-(d), the selected valence state histograms for the $V/V_0$ = 1.00, 0.87, 0.75, and 0.65 cases are visualized. Clearly, under ambient pressure (i.e., $V/V_0 = 1.00$), the $|N = 1,J = 2.5 \rangle$ atomic state is overwhelmingly dominant, which accounts for about 92.1\%. The probabilities for the other atomic states are small. As the pressure is increased, though the $|N = 1,J = 2.5 \rangle$ atomic state is still dominant, it is less prominent. The other atomic states start to play important roles, and their probabilities increase quickly with respect to the pressure. For example, at $V/V_0 = 0.65$, the $|N = 1,J = 2.5 \rangle$ atomic state only accounts for 64.2\%, while the probabilities for the $|N = 0,J=0.0\rangle$ and $|N=1,J=3.5\rangle$ atomic states are 13.5\% and 14.9\% [please refer to Fig.~\ref{fig:tprob2}(a)], respectively, which can not be ignored any more. Given the good quantum number $N$, we further sum up the probabilities of the corresponding atomic states. The results for different volume ratios are visualized using pie diagrams and presented in Fig.~\ref{fig:tprob1}(e)-(h). We found that at low pressure, the ruling contribution is from the $N = 1$ atomic states (i.e., the $4f^1$ electronic configuration). The proportions for the $N = 0$ and $N = 2$ atomic states (equivalent to the $4f^{0}$ and $4f^{2}$ electronic configurations) are quite trivial. As the pressure is increased the proportion of the $N = 1$ atomic states decreases while those of the $N = 0$ and $N = 2$ atomic states expand dramatically, i.e, a disproportionation-like process for the Ce-4$f$ electrons occurs. The contributions from the $N \geq 3$ atomic states are ignorable all the time. So, the overall trend is that the pressure will strengthen the 4$f$ valence fluctuation and enhance the mixed-valence behavior in CeIn$_{3}$ greatly.

The Ce-$4f$ valence fluctuation in CeIn$_{3}$ under pressure will result in extremely rich consequences. In Fig.~\ref{fig:tprob2}, we illustrate the evolution of atomic state probabilities for the three principal atomic states (including the $| N = 0, J= 0.0 \rangle$, $|N = 1, J=2.5\rangle$, and $|N = 1,J=3.5\rangle$ states), the averaged 4$f$ occupancy $\langle N_{f}\rangle$, and the averaged total angular momentum $\langle J \rangle$. The $\langle N_{f} \rangle$ and $\langle J \rangle$ were calculated using the following equations:
\begin{equation}
\langle N_{f}\rangle = \sum_{\Gamma} p_{\Gamma} N_{\Gamma}, 
\end{equation}
and
\begin{equation}
\langle J \rangle = \sum_{\Gamma} p_{\Gamma} J_{\Gamma}.
\end{equation}
Here $\Gamma$ denotes the index of atomic state, $p_{\Gamma}$ means the probability for the atomic state $|\Gamma\rangle$, $N_{\Gamma}$ and $J_{\Gamma}$ are the 4$f$ occupancy and total angular momentum for $|\Gamma \rangle$, respectively. As is seen in Fig.~\ref{fig:tprob2}(b) and (c), $\langle N_{f}\rangle$ and $\langle J \rangle$ decline with respect to the pressure, which are consistent with the previous discussion about the redistribution of the valence state histogram. The most noteworthy finding is that these two curves show obvious dips in the region of $V/V_0 \sim [0.80,0.90]$. Moreover, similar kinks are also identified around $V/V_0 \sim 0.85$ in the $p_{\Gamma}$ curves for the three principal atomic states [see Fig.~\ref{fig:tprob2}(a)]. We took the derivatives for them with respect to $V/V_0$, and observed a big ``hump" in the same region. Note that Shim \emph{et al.} also found similar results in the temperature-driven localized-itinerant electronic transition for another heavy fermion system CeIrIn$_{5}$~\cite{Shim1615}. Thus, we believe that these abnormal features are tightly connected with the change of the 4$f$ localized degree of freedom. Since we carried out DFT + DMFT calculations for CeIn$_{3}$ at high temperature, and we didn't observe any singularities (such as divergence, quick jumps or collapses) in the calculated quantities, it is more likely a crossover instead of a transition. In other words, these exotic features provide some useful fingerprints to characterize the pressure-tuned 4$f$ localized-itinerant crossover in CeIn$_{3}$ and we can use them to make a rough but meaningful estimation for the critical pressure $P_c$ or crystal volume $V_c$. According to the experimental and theoretical $P-V$ equation of states~\cite{PhysRevB.80.125131}, we speculate that the critical pressure at $T = 116$ K is about 7.0 $\sim$ 24.0 GPa. On the other hand, if we extrapolate linearly the available $P_c - T^{*}$ data~\cite{PhysRevB.77.064508} to $T^{*} = 116$ K, the corresponding $P_c$ is around 5.0 $\sim$ 15.0 GPa. Thus, further experiments are highly desired to judge which estimation is more reasonable. 

Besides $p_{\Gamma}$, $\langle N_f \rangle$, and $\langle J \rangle$, the pressure-driven 4$f$ localized-itinerant crossover and valence fluctuation in CeIn$_{3}$ also have considerable influences on the other physical observables. In Fig.~\ref{fig:tene}, we show evolution of the kinetic energy $E_{\text{kin}}$, potential energy $E_{\text{pot}}$, spectral weight of the $j = 5/2$ states in the vicinity of Fermi level $W_{j = 5/2}$, and the low-energy scattering rate $\gamma$ of the 4$f$ electrons. The definitions for $W_{j = 5/2}$ and $\gamma$ are expressed as follows:
\begin{equation}
\label{eq:w52}
W_{j = 5/2} = \int^{a}_{b} A^{j = 5/2}_{4f}(\omega) d\omega,
\end{equation}
and
\begin{equation}
\gamma = -\Im \Sigma_{j = 5/2} (\omega = 0),
\end{equation}
where $A^{j = 5/2}_{4f}(\omega)$ is the partial density of states for the Ce-4$f$ $j = 5/2$ states [see Fig.~\ref{fig:tdos}(b2)] obtained by DFT + DMFT calculations, and $[a,b]$ is the approximative energy range where the characteristic peak exists. Here we chose $a = -0.22$\ eV and $b = +0.18$\ eV for the $j = 5/2$ states. As for the $E_{\text{kin}}$, $E_{\text{pot}}$, and $W_{j = 5/2}$, they increase quickly with the increasing pressure, which means the enhancement of metallicity and itinerancy. However, when the crystal volume is shrunk, the low-energy scattering rate $\gamma$ approaches zero rapidly, which indicates the crossover from non-Fermi-liquid state to Landau Fermi-liquid state. All of these quantities show obvious ``kinks" or ``fluctuations" in the region of $V/V_0 \in [0.80,0.90]$ again. As discussed above, this volume range is probably related to the regime where the Ce-$4f$ localized-itinerant crossover would occur in CeIn$_{3}$. 

\section{concluding remarks\label{sec:summary}}

In the present work, we performed charge fully self-consistent DFT + DMFT calculations to study the pressure-driven 4$f$ localized-itinerant crossover in cubic CeIn$_{3}$ at finite temperature ($T \sim 116$ K). We mainly focused on the evolution of its electronic structures under pressure. The calculated results include the partial and total density of states $A_{4f}(\omega)$ and $A(\omega)$, momentum-resolved spectral functions $A(\mathbf{k},\omega)$, valence state histograms $p_{\Gamma}$, averaged 4$f$ occupancy $\langle N_f \rangle$, averaged total 4$f$ angular momentum $\langle J \rangle$, and low-energy scattering rate $\gamma$, etc. We found that upon increasing pressure, the spectral weights near the Fermi level, which are mainly associated with the $j = 5/2$ and $j= 7/2$ states, get bigger and bigger. On the other hand, the valence state histograms also exhibit great changes, such as the contribution from the $|N = 1, J = 2.5\rangle$ atomic state decreases, while those from the $|N = 1, J = 3.5\rangle$ and $|N = 0, J = 0.0 \rangle$ atomic states increase remarkably. Some calculated quantities, such as $p_{\Gamma}$, $\langle N_f \rangle$, $\langle J \rangle$, $E_{\text{kin}}$, $E_{\text{pot}}$, $W_{j = 5/2}$, and $\gamma$, etc., show abnormal behaviors in $V/V_0 \in [0.8,0.9]$, which are likely the signatures for the Ce-$4f$ localized-itinerant crossover. We can utilize these features to locate the critical pressure and temperature for the crossover.

The localized-itinerant transition (or crossover) and related valence fluctuation properties usually exist in many rare-earth heavy fermion systems~\cite{RevModPhys.56.755,coleman:2007}. There are still quite a lot of questions and puzzles to be answered and solved, such as whether the Fermi surfaces in YbRh$_{2}$Si$_{2}$ are temperature-dependent or temperature-independent~\cite{PhysRevX.5.011028}, and how the localization freedom of degree of 4$f$ electrons are affected by temperature, pressure, and external fields, etc. In order to reveal the rich physics underlying these complex phenomena, it is essential to employ the modern first-principles many-body approach (such as the DFT + DMFT method) which should consider the strong Coulomb interaction and the spin-orbit coupling on the same footing. The present work is probably the first systematically \emph{ab initio} investigation concerning the localized-itinerant crossover in the Kondo lattice compound CeIn$_{3}$. Though the spatial quantum fluctuation which may be essential to correctly capture the interplay between the 4$f$ localized and itinerant states is completely ignored in our DFT + DMFT calculations, the calculated results seem pretty good. To take the spatial quantum fluctuation into consideration, the extensions of DMFT such as extended DMFT (EDMFT) should be necessary~\cite{PhysRevLett.99.227204}. However, to our knowledge, it is impossible to perform charge fully self-consistent DFT + EDMFT calculations for realistic materials until now. Further works should be undertaken in this field. 

In the present work, we propose an efficient way which employs the combinations of $p_\Gamma$, $\langle N_f \rangle$, $\langle J \rangle$, etc., to detect the pressure-driven 4$f$ localized-itinerant crossover. It is highly promising to apply this method to study the similar transitions or crossovers in the other heavy fermion systems (such as Ce-, Sm-, and Yb-based mixed-valence materials) or even the actinide systems (such as the mysterious U-, Pu-, and Am-based strongly correlated materials). 


\begin{acknowledgments}
We acknowledge the fruitful discussions with Prof. Yifeng Yang, Prof. Hongming Weng, and Dr. Jianzhou Zhao. This work was supported by the Natural Science Foundation of China (Grant No.~11504340), the Foundation of President of China Academy of Engineering Physics (Grant No.~YZ2015012), and the Discipline Development Fund Project of Science and Technology on Surface Physics and Chemistry Laboratory (Grant No.~ZDXKFZ201502). The DFT + DMFT calculations were performed on the UniFr cluster (in Fribourg University, Switzerland) and Delta cluster (in the Institute of Physics, CAS, China).
\end{acknowledgments}


\bibliography{cein3}

\begin{thebibliography}{60}
\bibitem [{cfs()}]{cfs}%
  \BibitemOpen
  \href@noop {} {}\bibinfo {note} {In CeIn$_{3}$, the crystal-field splitting
  $\Delta_{\text{CF}}$ and spin-orbit splitting $\Delta_{\text{SO}}$ are 11.6
  meV and 280 meV~\cite{PhysRevB.56.1620}, respectively. So in the present
  work, we only considered the spin-orbit coupling, and ignored the
  crystal-field splitting.}\BibitemShut {Stop}%
\end{thebibliography}
\end{document}